\documentclass[preprint,12pt]{elsarticle}

\usepackage{graphicx,dcolumn,amsmath,amssymb,color,mathrsfs,titlesec,hyperref,enumerate,bm,subfigure,float,comment,lineno,todonotes,soul}

\biboptions{sort&compress}

\newcommand{\ft}{{\mathcal{F}}}

\renewcommand{\vec}[1]{\bm{#1}}

\emergencystretch=\maxdimen
\usepackage{hyperref}
\definecolor{redlinks}{rgb}{1,0,0}
\hypersetup{
  colorlinks=true,
}

\journal{Ultramicroscopy}

\begin{document}

\begin{frontmatter}
\title{Patterned Probes for High Precision 4D-STEM Bragg Measurements}

\author[Address01]{Steven E Zeltmann}
\ead{steven.zeltmann@berkeley.edu }

\author[Address02]{Alexander M\"uller}
\author[Address02]{Karen C Bustillo}
\author[Address02]{Benjamin Savitzky}
\author[Address02]{Lauren Hughes}

\author[Address01,Address02]{Andrew M Minor}
\author[Address02]{Colin Ophus}
\ead{cophus@gmail.com}

\address[Address01]{Department of Materials Science and Engineering, University of California, Berkeley, Berkeley, USA, 94720.}
\address[Address02]{National Center for Electron Microscopy, Molecular Foundry, Lawrence Berkeley National Laboratory, Berkeley, USA, 94720.}

\begin{abstract}
Nanoscale strain mapping by four-dimensional scanning transmission electron microscopy (4D-STEM) relies on determining the precise locations of Bragg-scattered electrons in a sequence of diffraction patterns, a task which is complicated by dynamical scattering, inelastic scattering, and shot noise. These features hinder accurate automated computational detection and position measurement of the diffracted disks, limiting the precision of measurements of local deformation. Here, we investigate the use of patterned probes to improve the precision of strain mapping. We imprint a ``bullseye'' pattern onto the probe, by using a binary mask in the probe-forming aperture, to improve the robustness of the peak finding algorithm to intensity modulations inside the diffracted disks.  We show that this imprinting leads to substantially improved strain-mapping precision at the expense of a slight decrease in spatial resolution. In experiments on an unstrained silicon reference sample, we observe an improvement in strain measurement precision from 2.7\% of the reciprocal lattice vectors with standard probes to 0.3\% using bullseye probes for a thin sample, and an improvement from 4.7\% to 0.8\% for a thick sample. We also use multislice simulations to explore how sample thickness and electron dose limit the attainable accuracy and precision for 4D-STEM strain measurements.

\end{abstract}


\end{frontmatter}

\hypersetup{
    linkcolor=redlinks,
    urlcolor=redlinks,
    citecolor=redlinks
}

\section{Introduction}
\label{Sec:Intro}
Strain at the nanoscale is important in understanding deformation mechanisms of structural materials \cite{wang2012sample}, as well as for engineering of transport properties in semiconductor devices \cite{bedell2014strain}. Nanostructures can support strains of up to $\approx 10\%$ without relaxation, providing great opportunities to engineer properties in ways that are not available in bulk materials \cite{li2014elastic}. A variety of techniques exist for measuring deformation with nanometer-scale resolution, including X-ray ptychography \cite{holt2014xray} or coherent diffraction \cite{robinson2009coherent}, though at present the highest spatial resolution is achieved in the transmission electron microscope (TEM). TEM strain measurements have been accomplished by dark-field holography \cite{koch2010efficient,COOPER2016145}, atomic resolution imaging \cite{bierwolf1993direct,GALINDO20071186,HYTCH1998131}, and converged-beam techniques \cite{jones1977higher,zhang2006direct,clement2004strain,hytch2014observing}.

In scanning transmission electron microscopy (STEM), a converged electron probe is rastered across the sample, and some of the scattered electrons (usually those scattered incoherently by thermal diffuse scattering) are measured to assign a value to each pixel \cite{pennycook2011_STEM_textbook}. Modern electron detector technology allows the full scattering pattern at each STEM probe position to be recorded, an experiment referred to as four-dimensional scanning transmission electron microscopy (4D-STEM) \cite{ophus_2019}. This method, also referred to as scanning electron nanodiffraction (SEND) or nanobeam electron diffraction (NBED), has been used in analyses of crystal orientation \cite{rauch2005rapid,brunetti2011confirmation,panova2016orientation}, local ordering of glassy states \cite{liu2015interpretation}, sample thickness \cite{lebeau2010position,zhu2012bonding}, and other analyses as described in a recent review \cite{ophus_2019}.

\begin{figure}
    \centering
    \includegraphics[width=3.4in]{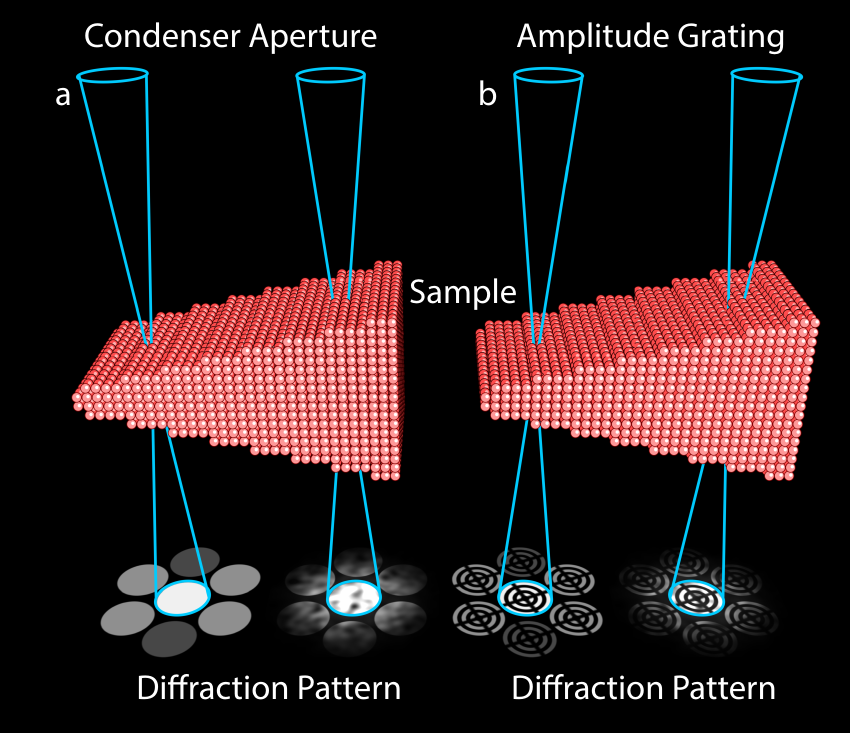}
    \caption{(a) Schematic of experimental setup for 4D-STEM strain mapping. A converged electron probe is rastered across the sample and a diffraction pattern is acquired at each probe position. Thick regions of the sample have complicated dynamical contrast inside the CBED disks that make accurate position determination difficult. In (b), a grating is inserted in the condenser system of the microscope to pattern the probe in momentum space. This pattern is imprinted on the diffracted disks, providing sharp edges in registry with the probe pattern that makes computational determination of their position more robust. }
    \label{fig:schematic}
\end{figure}

4D-STEM is used for mapping strain at the nano-scale by locating the Bragg scattered electrons in each pattern, whose position on the detector is related to the local lattice spacing. This approach has been used to map strain in electronic devices \cite{usuda2004strain}, structural materials \cite{pekin2017strain}, including \textit{in situ} deformed samples \cite{gammer2018local,pekin2018situ}, two-dimensional materials \cite{han2018strain}, and other systems where nanoscale deformation is of interest. 4D-STEM allows a large field of view and flexibility with regards to sample type and orientation \cite{beche2009improved,rouviere2013method}. Figure~\ref{fig:schematic}a shows a schematic view of the experimental setup for 4D-STEM strain mapping---the convergence angle is chosen so that non-overlapping convergent beam electron diffraction (CBED) disks are obtained in each pattern. 

In investigations of mechanical deformation and strain-engineered semiconductor devices the strains of interest are generally on the order of $\approx 1\%$, which is much larger than the currently achievable precision, reported to be  $6\times10^{-4}$ \cite{guzzinati2019electron} using the standard microprobe-STEM mode (i.e. without precession or patterned probes). This precision is not sufficient for several potential applications of 4D-STEM strain mapping, such as temperature mapping by thermal expansion measurement or mapping certain structural transformations via the lattice parameters, where strains may be on the order of $10^{-4}$. We note that direct comparison between precision limits reported in the literature is difficult because the precision limit depends on the sample properties, microscope image distortions, and the electron dose \cite{mahr2015theoretical, pekin2017strain, grieb2017optimization}.

The precision of the strain maps obtained by 4D-STEM is governed by the precision with which the Bragg scattered electrons can be located in each diffraction pattern. Non-uniform intensity of the diffracted disks, which can be caused by sample bending or dynamical diffraction in thick samples \cite{cowley1957scattering}, makes accurate detection of the positions of the diffraction disks difficult. Reducing the convergence angle of the electron probe shrinks the diffraction disks, hiding some of the dynamical effects at the expense of a larger real-space probe size. For this reason, much of the existing literature on 4D-STEM strain mapping uses convergence angles 0.2--1~mrad.
When operating at larger convergence angles, the centers of mass of the diffracted disks are not necessarily at the reciprocal lattice points, thereby requiring methods sensitive to the locations of the edges of the disk \cite{rouviere2013improved,pekin2017strain}. Disk position detection is often accomplished by cross- or phase-correlation of the diffraction pattern with a template image. These methods are still not ideal, as simulations performed by Mahr \textit{et al} \cite{mahr2015theoretical} found that the inner structure of the CBED disks is the limiting factor for precision of 4D-STEM strain measurements.

Post-processing of the 4D-STEM data and sophisticated data analysis methods have been shown to improve the precision of strain measurements. Pekin \textit{et al} \cite{pekin2017strain} investigated the optimal image filtering and correlation algorithms for diffraction disk detection, as well the robustness to non-uniform diffracted disks and signal-to-noise level. They found that the precision of disk location measurements can be degraded by an order of magnitude due to uneven illumination of the CBED disks. More computationally intensive disk-finding algorithms have also been implemented \cite{mahr2015theoretical,muller2012scanning}.

Changes to the experimental setup provide another route to improve precision. Precession of the incident electron beam with incoherent summation of the diffraction patterns at each beam tilt ``averages out'' dynamical contrast and illuminates higher-order diffraction disks, which can yield a substantial improvement in strain precision to $2\times10^{-4}$ \cite{rouviere2013improved}. However, this procedure requires specialized hardware in order to precess the beam in combination with scanning, and longer acquisition times. Mahr \textit{et al} \cite{mahr2015theoretical} showed simulations of the precision of 4D-STEM strain measurements for different experimental conditions, and suggested the use of patterned probes, but found no substantial improvement over standard circular apertures when imprinting a single cross on the probe. ``Hollow-cone'' or Bessel structured probes, produced using an annular condenser aperture, are akin to precession diffraction, but with all tilts illuminated simultaneously (and thus added coherently). Such probes were simulated and realized experimentally by Guzzinati \textit{et al} \cite{guzzinati2019electron}, yielding strain precision of $2.5\times10^{-4}$, rivaling precession diffraction. This approach also allows for higher convergence angles, as the sparsity of the patterned probe reduces the interference between the scattered beams. Diffraction patterns through thick samples also contain a large background intensity due to inelastic scattering, which can be effectively eliminated by zero-loss energy filtering \cite{hahnel2012improving,wehmeyer2018measuring}.

In this paper, we investigate the use of probes with patterning in momentum space to improve the robustness of cross-correlation disk detection. Using an amplitude grating in the probe-forming aperture of the condenser system imprints known patterning on the diffraction pattern that allows accurate position location even in the presence of highly non-uniform illumination of the diffracted disks, as shown schematically in Figure~\ref{fig:schematic}b. Such patterned apertures are easily fabricated by physical vapor deposition and focused ion beam (FIB) machining; are mechanically stable; and, due to high conductivity, do not suffer from charging artifacts. We used mutlislice simulations to optimize the design, and estimate the improvement in accuracy and precision of disk detection for patterned apertures relative to typical circular probes. We also carried out 4D-STEM strain measurement experiments on unstrained silicon samples and characterized the improvement of precision when using patterned probes.

\section{Theory}
\label{Sec:Theory}

\subsection{Measuring Disk Positions}

\begin{figure}
    \centering
    \includegraphics[width=3.2in]{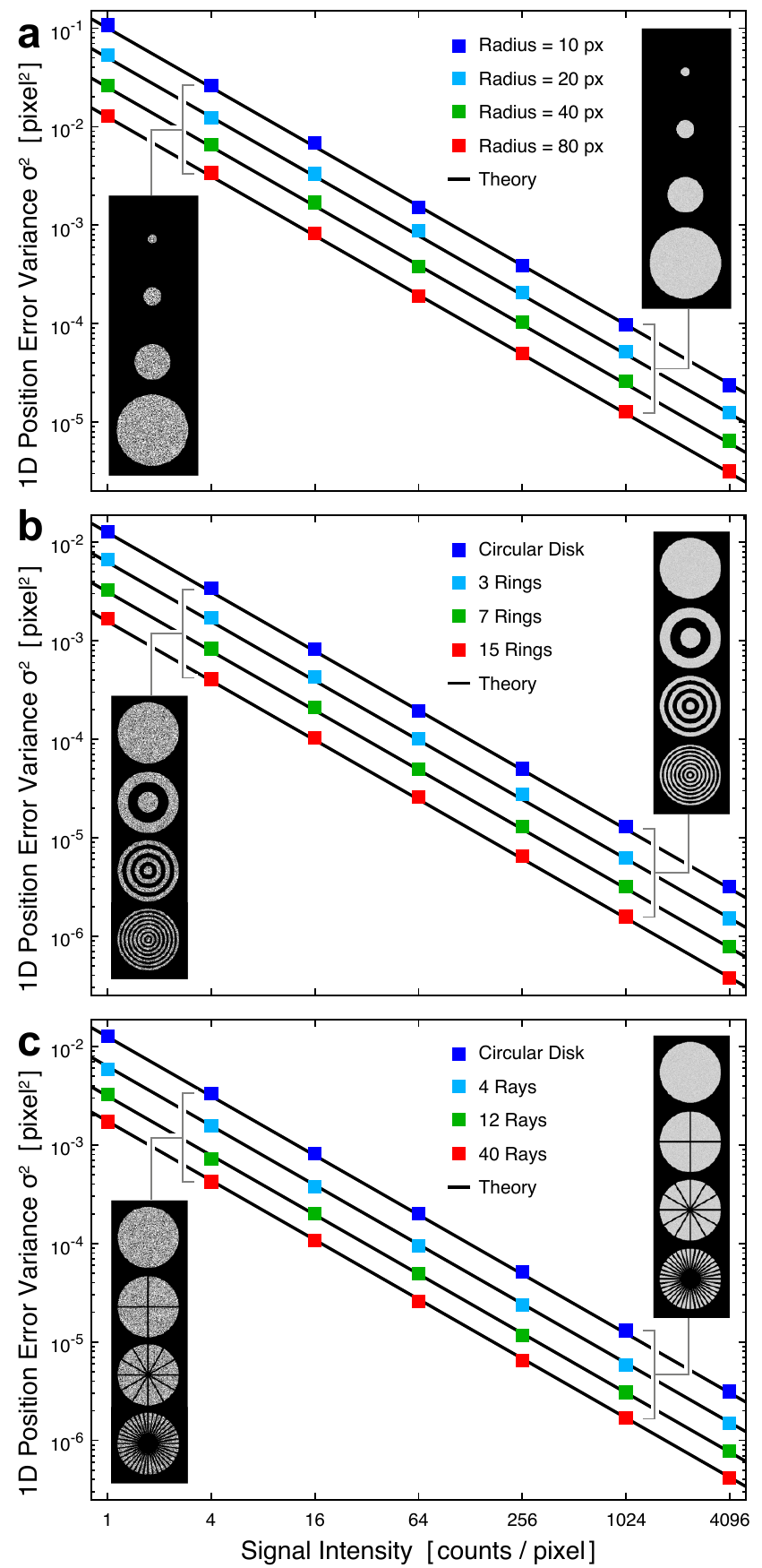}
    \caption{Numerical tests of image registration of an ideal STEM probe with a noisy measurement. Position error was measured for 1000 randomly generated probes along one dimension, for (a) circular disks with different radii, (b) varying numbers of concentric rings, and (c) varying numbers of intersecting rays. These measurements are compared to the theoretical precision given by Eq.~\ref{eqn:variance1}. Inset images show examples of noisy measurements.}
    \label{fig:TheoryDiskPrec}
\end{figure}

We determine the position of both scattered and unscattered Bragg disks by measuring the relative translation between a template image $I_{\rm{ref}}(\vec{r})$ and a disk image $I(\vec{r})$ using digital image correlation. This correlation image $I_{\rm{corr}}(\vec{r})$ can be determined efficiently by taking the Fourier transforms $\ft\{\}$ of each image,
\begin{eqnarray}
    G(\vec{q}) & \equiv & \ft \{ I(\vec{r}) \}
    \nonumber \\
    G_{\rm{ref}}(\vec{q}) & \equiv & \ft\{ I_{\rm{ref}}(\vec{r}) \} 
    \nonumber
\end{eqnarray}
and then using the expression,
\begin{eqnarray}
    I_{\rm{corr}}(\vec{r}) &=& 
        I(\vec{r}) \otimes I_{\rm{ref}}(\vec{r}) \\
        &=& \ft^{-1} \left\{ \frac{ 
        G(\vec{q}) G_{\rm{ref}}^*(\vec{q}) }
        {| G(\vec{q}) G_{\rm{ref}}^*(\vec{q}) |^p} 
         \right\}, \label{eqn:corr} 
\end{eqnarray}
where $\vec{r}=(x,y)$ and $\vec{q}=(q_x,q_y)$ represent the real space and reciprocal space coordinates respectively, $\otimes$ is the correlation operator, $^*$ indicates the complex conjugate, and $p$ is the correlation power law coefficient. The cross-correlation is given when $p=0$, and phase correlation is defined by $p=1$. Values of $p$ between 0 and 1 define a hybrid image correlation \cite{pekin2017strain}. In this work, we use cross-correlation with $p=0$ for all simulations, and both cross and hybrid ($p=0.25$) for the experimental data.

To estimate the error of a measured disk position, we follow the methods of Clement \textit{et al} \cite{clement2018image}. We first assume an ideal, noise-free measurement of the template probe image $I_{\rm{ref}}(\vec{r})$ is available, from careful measurements of the vacuum probe image. Next, we assume the measured image of a disk $I(\vec{r})$ has a signal given by a Poisson distribution with a mean of $n$ counts per pixel, and therefore also a variance of $n$. The variance ${\sigma_x}^2$ of a cross-correlation measurement of the image translation error along the $x$ direction is given by
\begin{equation}
    {\sigma_x}^2 = \frac{1}{n D_x},
    \label{eqn:variance1}
\end{equation}
where $D_x$ is the normalized ``image roughness'' \cite{clement2018image} along the $x$ direction, given by
\begin{equation}
    D_x = \frac{1}{L_x L_y} \sum_{q_x,q_y}
    (2 \pi q_x)^2
    | G(\vec{q}) |^2
\end{equation}
where $L_x$ and $L_y$ are the image dimensions. This expresses that the addition of more edges to the image template will lead to greater precision, as the presence of more edges will weight the higher Fourier coefficients more heavily. In addition, upsampling a band-limited image will increase the image dimensions $L_x, L_y$ without increasing the higher Fourier components and lead to decreased precision. 

If all units are in pixels, the image roughness for a circular disk with radius $R$ is given by $D_x \approx R$. Using this expression in Eq.~\ref{eqn:variance1} gives a variance of
\begin{equation}
    \left. {\sigma_x}^2 = \frac{1}{n R} \right. .
    \label{eqn:variance2}
\end{equation}
Note that this expression will often have a small numerical prefactor $\approx 1$ due to image details such as the maximum bandwidth and sharpness of the edges. The 2D variance will be given by ${\sigma_x}^2 + {\sigma_y}^2$.  To verify the above analysis, we performed numerical measurements of the disk position error for circular disks with various radii and counts per pixel. These measurements are shown in Figure \ref{fig:TheoryDiskPrec}a, and are in excellent agreement with Eq.~\ref{eqn:variance2}. 

To lower the disk position  error, we must increase the image roughness $D_x$ and $D_y$. One possibility is to add a series of concentric rings, as in Figure \ref{fig:TheoryDiskPrec}b. For $M$ total concentric rings that are linearly spaced, the image roughness is given by
\begin{equation}
    \left. {\sigma_x}^2 = \frac{2}{n R (M + 1)} \right. .
    \label{eqn:variance-bullseye}
\end{equation}
Increasing the number of concentric rings to 3, 7, or 15 will decrease the disk position error variance by factors of 1/2, 1/4, and 1/8, respectively.

An alternative method of increasing the image roughness $D_x$ and $D_y$ is to add linear ray features, radiating out from the center of the disk as shown in Figure \ref{fig:TheoryDiskPrec}c. An increasing number of rays lowers the position error variance, by increasing the image roughness $D_x$. Interestingly, combining concentric rings with linear rays does not further decrease the position error, though it can, in some circumstances, reduce the total number of counts while maintaining the same position error variance.

Finally, we note that the position error of a circular disk in terms of the total electron dose $N = \pi R^2 n$ is
\begin{equation}
    \left. {\sigma_x}^2 = \frac{\pi R}{N} \right. .
    \label{eqn:variance3}
\end{equation}
Thus we see that for a constant disk radius $R$, the variance has the expected scaling of $1/N$.  For a constant electron dose $N$, the variance scales linearly with radius $R$. This represents the fundamental trade-off between real space and reciprocal space error for Bragg disk position measurements. Increasing the probe's outer angle will generate a smaller probe in real space and thus improve real space resolution, but will worsen the measurement precision in reciprocal space. 

\begin{figure}
    \centering
    \includegraphics[width=3.5in]{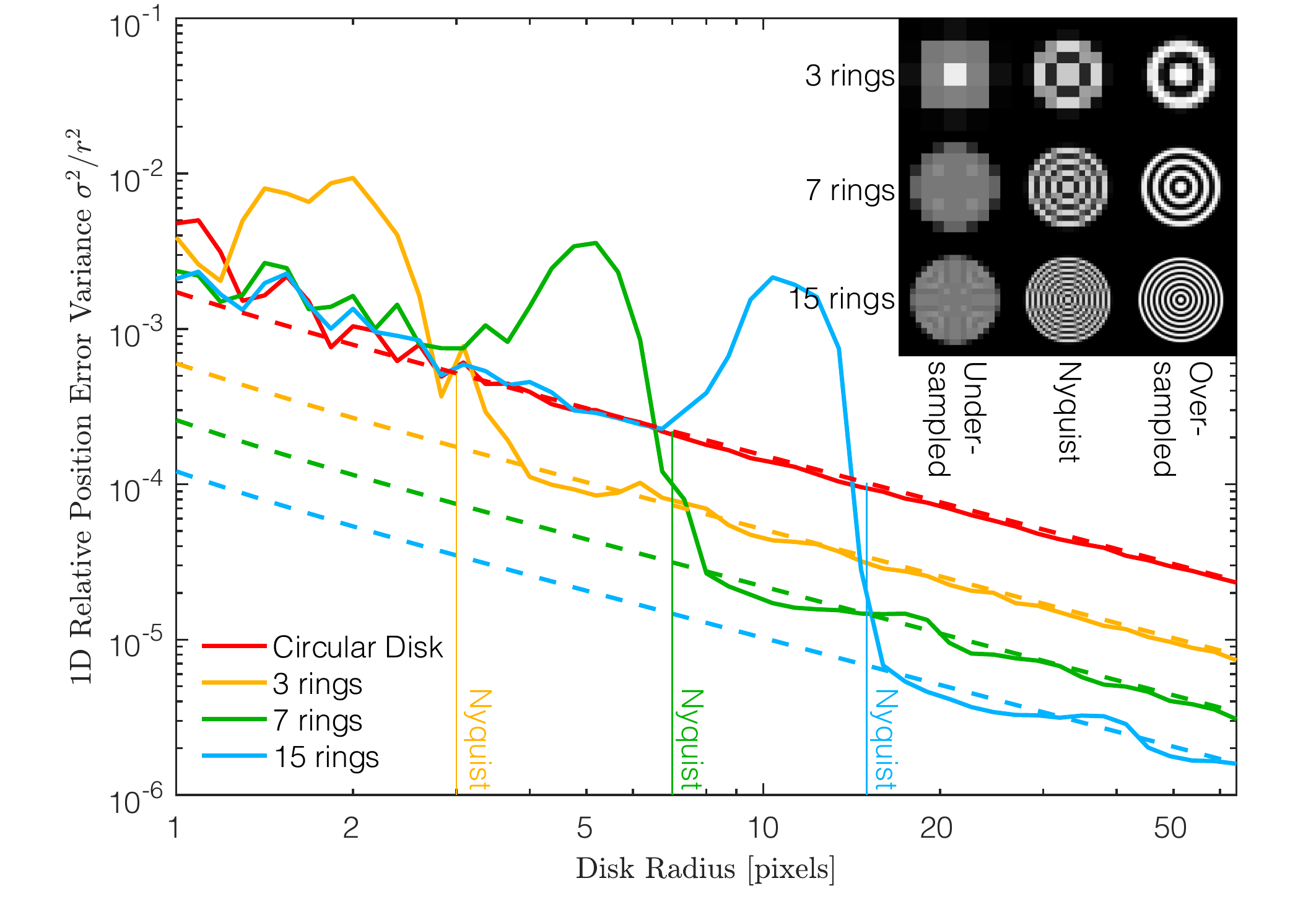}
    \caption{Numerical tests of the impact of sampling on relative position error variance. Concentric ring probes show the scaling expected from the theory at greater than Nyquist sampling, but have substantial position error due to aliasing at lower sampling. Dashed lines are the prediction of Equation~\ref{eqn:variance-bullseye}. Inset: example images of under-, Nyquist, and over-sampled disk images for 3, 7, and 15 ring disks.} 
    \label{fig:bullseye-pixsize}
\end{figure}{}

In order to realize the benefits of the patterned probes on the disk detection precision, the imprinted features inside the disks must be sufficiently resolved by the detector. The effect of the detector resolution is shown in Figure~\ref{fig:bullseye-pixsize} for probes with varying number of concentric rings and a constant dose of 1024 \textit{total} counts. Note that the lines on the plot for the theory drop the factor of 2 in Eq.~\ref{eqn:variance-bullseye}, which arose from the ``missing'' pixels cut off by the pattern, whereas here we fix the total dose such that the intensity per pixel roughly doubles inside the illuminated portion of the pattern. Nyquist sampling of the patterned probes requires one pixel per ring (as marked on the figure). We observe that at slightly below Nyquist sampling the patterned probes show substantially worse performance as compared to even an unpatterned probe, and as the pattern collapses into only a few pixels it shows the same performance as an unpatterned probe. Thus while sampling of just above one pixel per ring is sufficient, it is preferable to oversample the pattern to avoid the catastrophic drop-off in precision at just below Nyquist sampling.

The above analysis for ideal disk position measurement will often underestimate the potential gains of using patterned probes because real experiments often contain a significant amount of background signal and fine structure imparted to the disks by dynamical diffraction. In the following sections, we will show how adding various amplitude features to the STEM probe can reduce the disk position error for both multislice STEM image simulations and STEM experiments.

\subsection{Probe Size}

\begin{figure}
    \centering
    \includegraphics[width=3.2in]{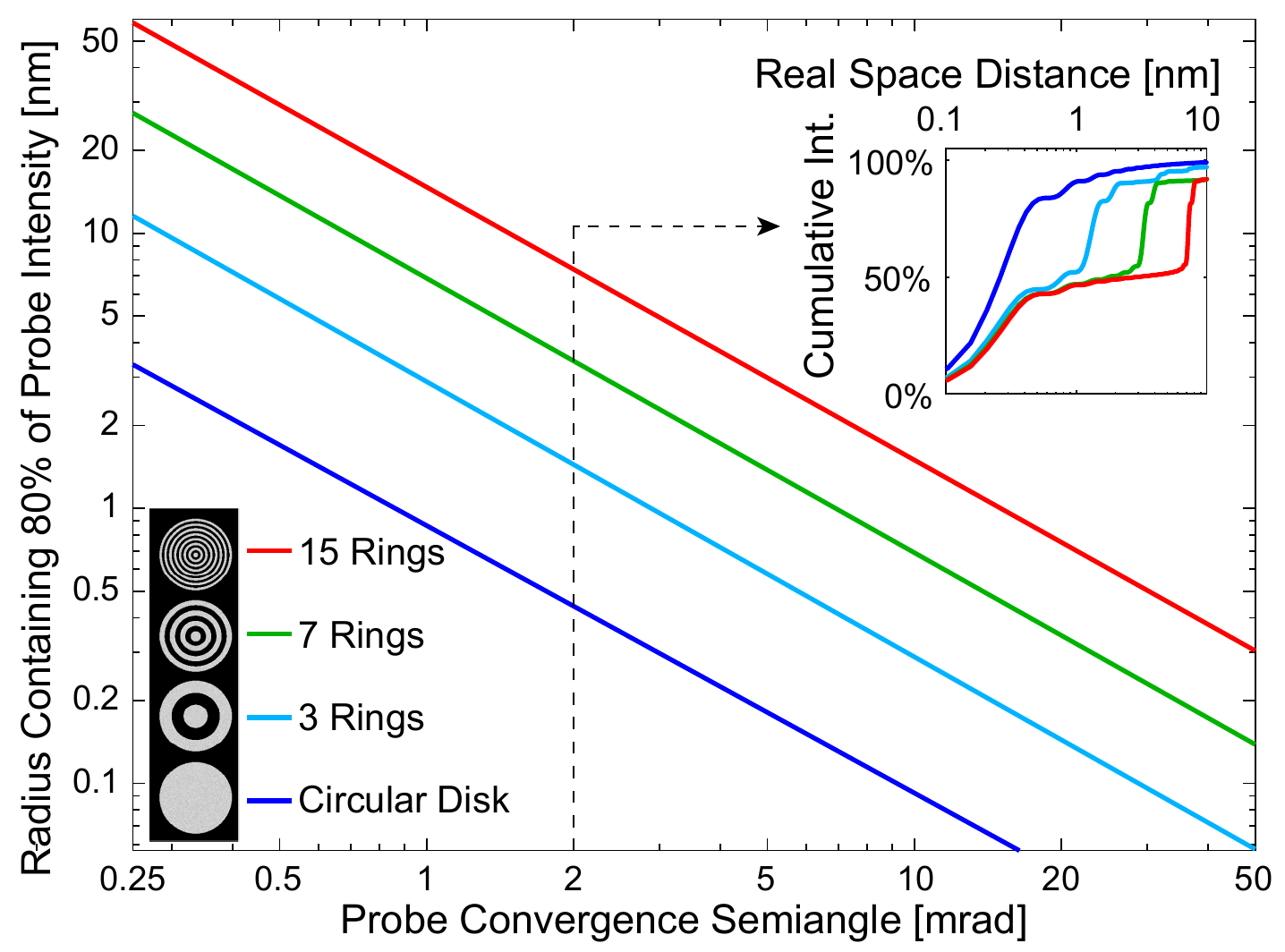}
    \caption{Numerical tests of the STEM probe size using different ring patterns at different convergence angles, for an aberration-free microscope at 300~kV. Adding the ring pattern to the probe causes the real-space probe size to grow by a factor determined by the number of rings. Inset: cumulative radial intensity profiles of probe intensity for a 2~mrad convergence semiangle.}
    \label{fig:probeSize}
\end{figure}

In STEM imaging experiments, especially at atomic resolution, it is usually advantageous to form as small a converged probe as possible. This is achieved when an aberration-free (flat phase) plane wave illuminates the probe-forming aperture, with as large a semi-convergence angle as possible. However, in 4D-STEM strain measurements, the minimum probe size should be the dimensions of the crystalline unit cell being measured. Increasing the probe size can be achieved by reducing the semi-convergence angle, or by adding amplitude patterns to the probe as described in the previous section. The dependence of the real space probe size on the number of patterned rings added to an aberration-free STEM probe is shown in Figure~\ref{fig:probeSize}. In order to include the effects of both increasing the size of the central lobe and increasing the intensity of the probe tails, we have defined the STEM probe size as the radius containing 80\% of the total probe intensity. The inset of Figure~\ref{fig:probeSize} shows the cumulative radial intensity of different patterned STEM probes with a semiconvergence angle of 2~mrad at 300~kV. The patterned probes have long tails that extend out from the center, decreasing the realspace resolution. Compared to the typically reported full width at half maximum, this probe size metric will overestimate the size of the probe, but better capture the effect of the long tails of the structured probes.

All STEM probe patterns lead to the same scaling law for probe size, where the probe size varies inversely with the semi-convergence angle. Adding additional amplitude rings to the probe will increase the prefactor of these power laws.  For example, Figure~\ref{fig:probeSize} shows that forming a 1 nm radius probe using 1, 3, 7, and 15 rings would require semi-convergence angles of 0.9, 3, 7, and 14 mrads respectively. Thus, when using patterned STEM probes, we will generally need to use somewhat larger semi-convergence angles to produce probes of the same size. When estimating the probe size in an experiment, the best practice is always to record real space images of the STEM probe in order to obtain an accurate estimate of the probe size and thus the spatial resolution.

\section{Methods}
\subsection{Multislice Simulations}

In 4D-STEM experiments, the strain mapping precision is not only dependent on single-disk matching precision, but rather on the precision of the lattice fit to several diffracted disks in a whole (near) zone axis pattern. To investigate the strain mapping precision taking account of the whole pattern fitting, we performed multislice simulations on an unstrained Si $\langle110\rangle$ model using a custom MATLAB code and potentials from Kirkland's parametrization \cite{kirkland2010advanced}, with some implementation details give in \cite{ophus2017fastSim}. Poisson random noise was applied to the diffraction patterns to simulate shot noise for different numbers of electrons per diffraction pattern. We simulated a 5~nm thick model to obtain diffraction patterns with largely kinematical scattering, and a 20~nm thick model to obtain patterns with dynamical contrast in the CBED disks. The convergence angle was chosen to be 2.7~mrad at 300~kV to provide nearly-touching CBED disks, which maximizes the real-space resolution while avoiding interference between the diffracted beams, and gives the worst-case scenario for disk location; the simulations are aberration-free, which also maximizes the local variations in the CBED disks. The bullseye pattern is rotated by an arbitrary amount to prevent aliasing artifacts that may arise if the bars with the simulation grid.

\subsection{Numerical Measurement of Disk Positions}

The procedure for obtaining strain maps from a 4D-STEM dataset involves (a) precisely locating the diffracted disks in each diffraction pattern, (b) obtaining a guess of the reciprocal lattice vectors $u_0$ and $v_0$, (c) using the approximate reciprocal lattice vectors to index each diffracted disk, and finally (d) solving an (overdetermined) linear least squares problem to obtain the best-fit $u$ and $v$ vectors for each diffraction pattern, from which we calculate the strain. All of the analysis in this work was performed using the open-source py4DSTEM Python module \cite{savitzky} available at the \href{https://github.com/py4dstem/py4DSTEM}{py4DSTEM github repository}.

In each diffraction pattern we locate the diffracted disks by taking the Fourier correlation (Eq.~\ref{eqn:corr}) of a convolution kernel, or template image, with each diffraction pattern. For experimental data, the convolution kernel can be obtained either by imaging the probe in diffraction through vacuum or by averaging the direct beam from many diffraction patterns. For simulated data, we use the initial wavefunction as the convolution kernel. The peaks in the correlation image between the kernel and the diffraction pattern correspond to the locations of the diffracted disks. The positions of the diffracted disks are further refined by subpixel registration using the matrix-multiplication discrete Fourier transform upsampling approach \cite{Guizar-Sicairos:08,Soummer:07} and a final local parabolic fitting \cite{gleason1991subpixel}. This subpixel refinement method locally upsamples the correlation image in a 1.5~px wide window around each correlation peak by a given factor (16 in this work), without computing the entire upsampled correlation image.
Each identified peak is indexed based on an initial guess of the lattice vectors, and linear least squares fitting is used to determine the reciprocal lattice vectors in each diffraction pattern. Each indexed peak is weighted by the correlation intensity in the least squares fit. Strain maps are then obtained by mapping the change in the lattice vectors.
There are several thresholds and filters applied in this procedure---while we slightly tune these parameters for the different simulated models and experimental samples, in all cases the normal probe and bullseye probe at each condition are processed with identical parameters.

\subsection{Cross-validation}
In measuring the strain mapping precision from the simulated data, we make use of ground truth knowledge of the sample, i.e. that the model was completely strain-free and there were no projection distortions. For real experimental data, there are artifacts that complicate this analysis: the sample may be bent or strained due to fabrication artifacts or beam heating, and the microscope projection system introduces astigmatism that distorts the pattern. Since strain information in 4D-STEM is calculated from the lattice fitted to the diffracted disks in each diffraction pattern, we can estimate the precision of the strain measurement by evaluating the agreement between the fitted lattice and the individual disk position measurements. 

While the residual error from the linear least squares fit of the lattice vectors is one such metric, because of the limited number of diffraction patterns in a dataset and the effects of the artifacts described above, for the experimental diffraction patterns we calculate a ``cross-validation'' error. Cross-validation is often used to evaluate the quality of high-dimensional models \cite{arlot2010survey}. In each diffraction pattern, half of the identified disks are chosen at random and a best-fit lattice is obtained from only these disks. The expected positions of the other half of the disks in the same pattern are computed from this lattice, and we define the error as the root mean square (RMS) difference between these predicted positions and the actual measured disk positions. For each diffraction pattern, we repeat this procedure of training on a random subset and testing against the other measurements 200 times per diffraction pattern to ensure statistical relevance.

\label{Sec:Methods}

\subsection{Bullseye Aperture Fabrication}
\label{ssec:fabrication}
We fabricated a set of bullseye apertures by FIB milling a gold-coated silicon nitride TEM window. Although the theory indicates adding linear rays to the concentric ring pattern does not improve strain precision, we included four rays for structural support in the fabricated apertures. An approximately 1~$\mu$m thick layer of gold was thermally evaporated onto the flat side of a 200~nm thick silicon nitride TEM window (Norcada, Canada) with a single 250~$\mu$m square window. Approximately one gram of gold was evaporated at a pressure better than $2 \times 10^{-6}$~torr, with the substrate kept at room temperature.  

\begin{figure}
    \centering
    \includegraphics[width=0.5\textwidth]{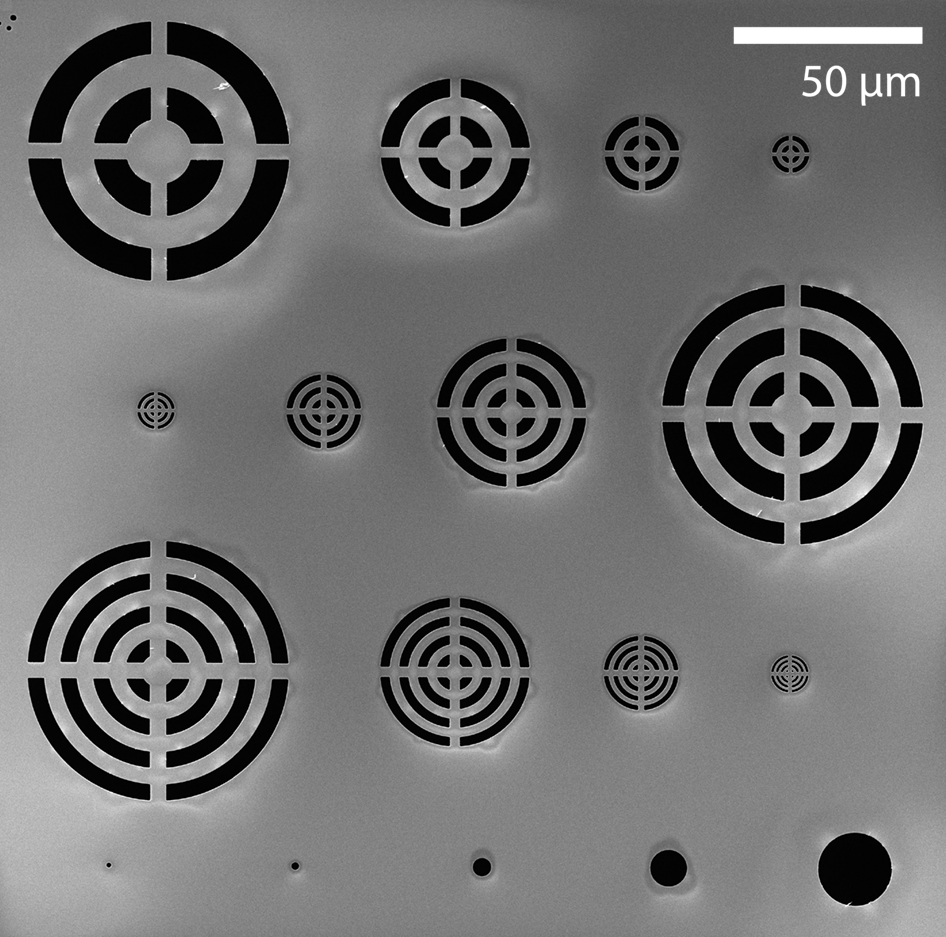}
    \caption{SEM micrograph of the fabricated bullseye aperture plate. }
    \label{fig:SEM-overview}
\end{figure}

The bullseye apertures were milled into the gold-coated window using a FEI Helios G4 UX dual beam SEM/FIB at 30~kV. The milled aperture plate is shown in Figure~\ref{fig:SEM-overview}. We milled bullseye patterns with 2, 3, and 4 rings and with 70, 40, 20, and 10~$\mu$m diameters. The 70, 40, and 10~$\mu$m bullseyes match the sizes of the standard circular apertures installed in our microscope, which simplifies beam alignments. In addition, we milled a set of circular apertures of 20, 10, 5, 2, and 1~$\mu$m diameter, which can be used to produce STEM probes with very small convergence angles or low beam current for imaging very dose-sensitive materials. Since the apertures are more closely spaced than is typical, electrons pass through all of them and a third condenser beam-forming aperture was therefore used to isolate a single probe for nanodiffraction experiments. 

\subsection{Strain Map Acquisition}
\label{ssec:methods-4dstem}
The bullseye aperture plate was installed in the second condenser aperture holder of a FEI TitanX operated at 300~kV. A silicon $\langle110\rangle$ sample was prepared by wedge polishing followed by Ar ion milling. 4D-STEM datasets were acquired with a scan size of 25$\times$25~pixels, diffraction pattern image size of 512$\times$512~pixels, and a probe semi-convergence angle of approximately 3~mrad. Diffraction patterns were acquired using a Gatan Orius 830 CCD. We obtained scanning diffraction datasets from two regions of the wedge sample: a ``thin'' region with relatively even illumination of the diffracted disks, and a ``thick'' region with substantial dynamical contrast.

\section{Results and Discussion}
\subsection{Multislice Simulations}

\begin{figure}
    \centering
    \includegraphics[width=\textwidth]{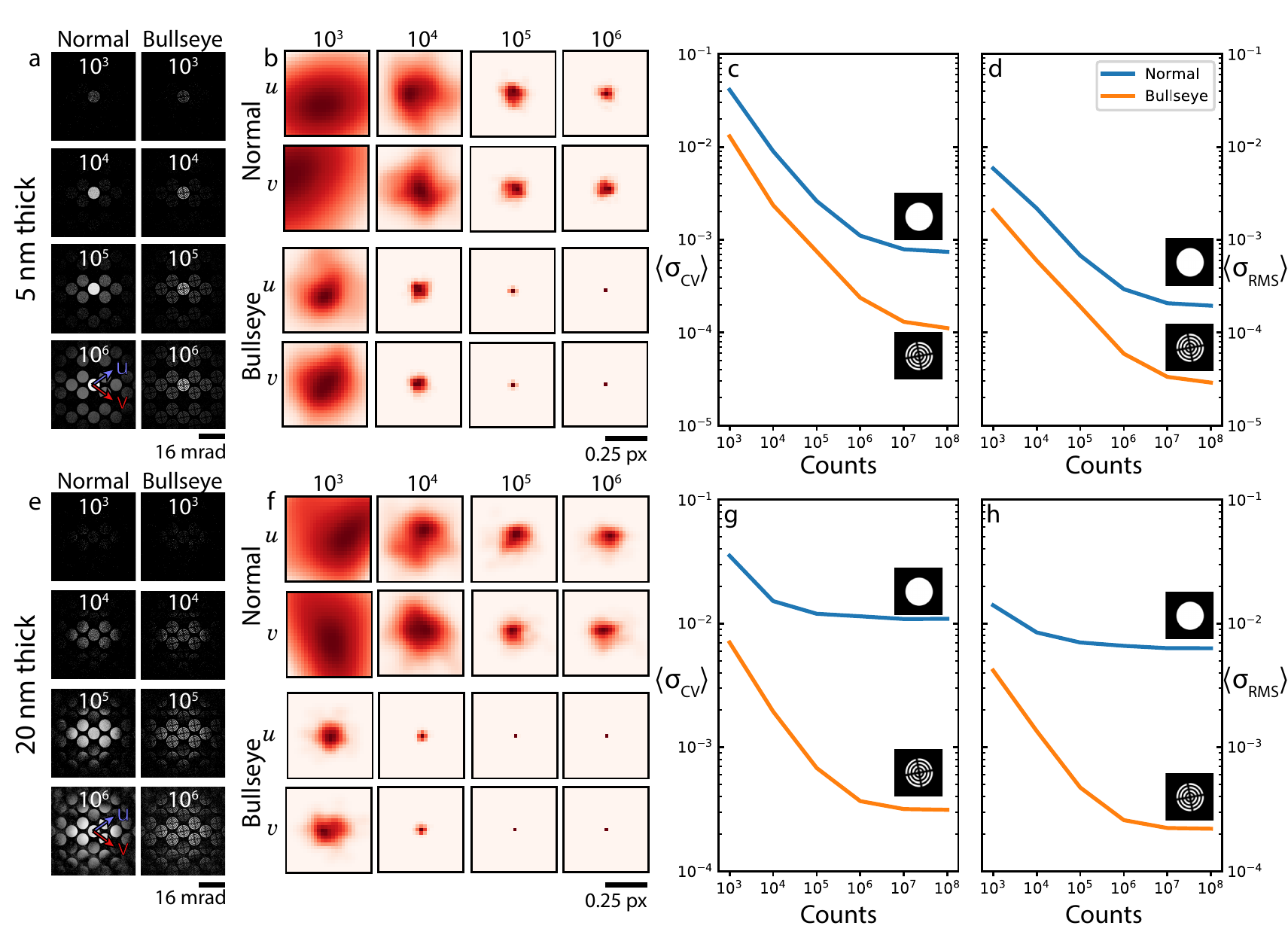}
    \caption{Strain mapping precision of simulated silicon diffraction data at different samples thickness and electron dose. (a,e) Representative simulated diffraction patterns at different electron doses per pattern. (b,f) Comparison of the $u$ and $v$ reciprocal lattice vectors measured at each scan position in the simulation of a strain-free sample. The center of each histogram represents the average $u$, $v$ positions obtained from the noise-free simulation data. (c,g) Cross-validation error and (d,h) RMS fit error, relative to the reciprocal lattice vector length (equivalent to the strain error in the small strain limit). (a-d) are obtained from a 5~nm model with largely kinematical scattering, while (e-h) are from a 20~nm model with dynamical contrast inside the CBED disks. The reciprocal lattice vectors are drawn in the bottom left panel of (a), and have length $\approx 70$~pixels.}
    \label{fig:multislice}
\end{figure}

Multislice simulations of 5 and 20 nm thick unstrained silicon along the $\langle 110 \rangle$ zone axis are shown in Fig.~\ref{fig:multislice}. The diffraction patterns in Fig.~\ref{fig:multislice}a from the 5~nm model show even illumination of the CBED disks and the (002) forbidden reflection is not excited. The diffraction patterns in Fig.~\ref{fig:multislice}e from the 20~nm model show uneven illumination of the disks and the (002) reflection is partially illuminated due to double diffraction. 

Figure~\ref{fig:multislice}b and f show the locations of the $u$ and $v$ reciprocal lattice vectors identified in each diffraction pattern of the simulated 4D-STEM scans, illustrating the variation in the measured lattice vectors as the probe scans across a totally strain-free sample. In the limit of small strains, the uncertainty in the reciprocal lattice vectors relative to the reciprocal lattice vector length is equal to the uncertainty in the measured strain. The center of each histogram corresponds to the lattice vectors measured from the 5~nm model with infinite dose, which we take as the ground truth. The $u$ and $v$ vectors correspond to the $(1\bar{1}1)$ and $(\bar{1}11)$ reflections (drawn in the bottom left panel of Fig.~\ref{fig:multislice}a) and each have a length of $\approx70$~pixels. Both the normal and bullseye probes converge to the same lattice vectors at high dose, though in all cases the spread of values is substantially larger for the normal probes. These wide variations in the lattice vectors from an unstrained sample lead to correspondingly large fluctuations in the calculated strain values. The asymmetric error in the histograms is likely due to the presence of the partially-illuminated forbidden reflections, which causes the cross-correlation peak uncertainty to be larger in one direction \cite{pekin2017strain}.

Figure~\ref{fig:multislice}d and h show the RMS residual error of the linear least squares lattice vector fitting, relative to the length of the $(\bar{1}11)$ reciprocal lattice vector. This error is one metric for the precision of the strain measurement as it reflects the uncertainty in the fitted lattice vectors. The error decreases with increasing electron dose and the bullseye apertures have $\approx3.5$~times lower error at up to $\approx10^5$~counts. For this number of rings, the image roughness metric (Eq.~\ref{eqn:variance-bullseye}) predicts a 4-fold increase in the precision of locating a single diffraction disk, without accounting for the presence of the incoherent background counts found in the multislice results.  As the illumination of each diffraction disk varies across the pattern, the location precision of each diffraction disk also varies, complicating comparison with the single-disk location precision theory. At higher signal levels, the error stops decreasing as we approach the limits of the subpixel fitting algorithm. At $10^8$~counts and above, the bullseye apertures show $\approx 7$~times improved precision.

When dynamical contrast causes intensity variations inside the diffracted disks, the normal probes show substantially worse performance. When locating the disks by cross-correlation, as in this calculation, the location assigned to each disk is biased towards the center of mass of the disk. In the simulated diffraction patterns, many disks are seen to be half-illuminated, which leads to substantial position errors regardless of the number of counts. The patterned probes are less sensitive to this type of error, as the cross-correlation intensity should peak when the rings are in registry even if the rings are not fully illuminated. Thus we observe in Fig~\ref{fig:multislice}g and h that the precision of the normal probe saturates by $10^5$ counts while the bullseye probe precision improves with increasing dose until $10^7$ counts. This robustness to uneven disk illumination gives the bullseye probes an even larger precision advantage compared to the kinematical case for thin specimens, with the minimum error decreasing by $\approx$~30 times at high electron counts.

\subsection{Experimental Measurements of Strain in Silicon}

\begin{figure}
    \centering
    \includegraphics[width=\textwidth]{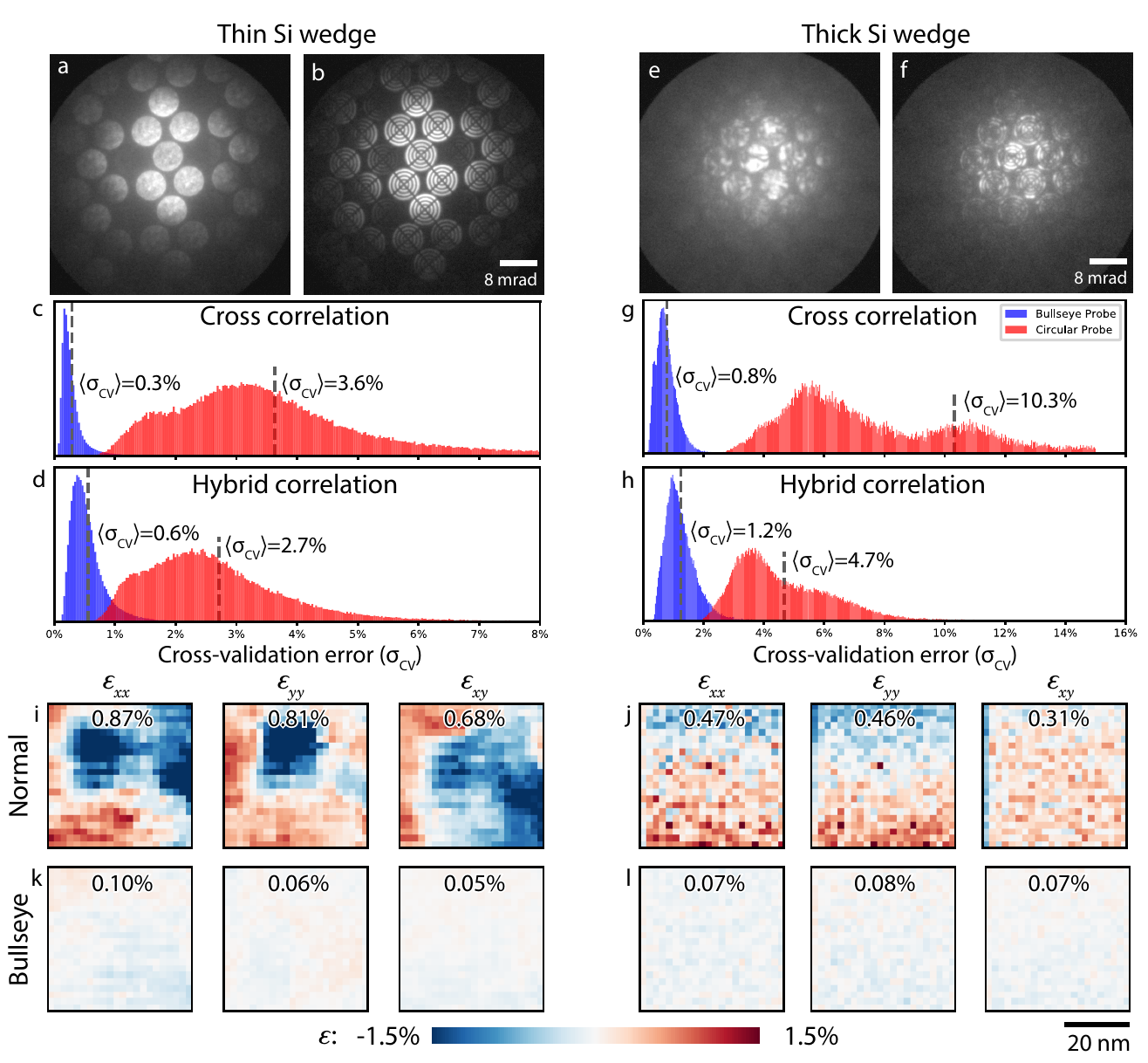}
    \caption{ Cross-validation error determination from 4D-STEM experiments on a Si $\langle 110 \rangle$ wedge. Diffraction patterns from a thin region of the wedge with (a) the standard circular aperture and (b) with the bullseye amplitude grating. The cross validation error, computed by fitting a lattice to half of the identified diffraction disks and measuring the error of the remaining half, using (c) cross-correlation and (d) hybrid fitting. Diffraction patterns from a thick region of the wedge (e) without and (f) with the bullseye aperture. Cross-validation strain error for (g) cross-correlation and (h) hybrid correlation disk detection. Strain maps from each region of the Si wedge sample are shown in (i)--(l). The label on each strain map incates the standard deviation of that strain component over the field of view.}
    \label{fig:experiment}
\end{figure}

Representative diffraction patterns from the scans are shown in the top row of Figure~\ref{fig:experiment}. In the thick scan region, the (002) forbidden reflection is fully illuminated and there is substantial dynamical contrast inside the CBED disks.

As we cannot guarantee that the silicon specimen is strain-free, we cannot use the spread in the measured lattice vectors as an indicator of the precision of the measurement, and instead report only the cross-validation error for the experimental scans. The cross-validation (CV) error relative to the length of the (111) reciprocal lattice vector for the experimental scans is shown in Figure~\ref{fig:experiment}c and g. When finding the disk locations in the thin region by cross-correlation, use of the bullseye patterned probe causes the mean CV error score to decrease to 0.3\% from 3.6\%, an improvement of 12 times. In the thick region, the CV score decreases to 0.8\% from 10.3\%, an improvement of $\approx13$~times. The improvement in strain precision we observe in experiments is larger than predicted by the image roughness theory and observed in the multislice results. The inelastic component present in the experimental data likely plays a large part in this discrepancy, as the presence of substantial intensity between the Bragg disks reduces the contrast between the disks and the background, compounding the center-of-mass bias in the unpatterned probes and increasing the impact of the patterned bars in the disks on matching precision.

While cross-correlation performs well when the pattern background is low, hybrid correlation has been shown to better handle the `plasmonic blur' in real samples \cite{pekin2017strain}. To test if hybrid correlation can also improve the strain precision when using patterned probes we repeated the disk finding procedure with $p=0.25$ (Eq.~\ref{eqn:corr}). We observe that the CV error  decreased substantially compared to cross-correlation for the normal probes, as shown in Figure~\ref{fig:experiment}d and h. However, in both the thick and thin Si wedge regions, the CV error of the bullseye apertures was 50--100\% worse when using hybrid correlation. The image roughness theory discussed above does not generalize easily to hybrid correlation and does not account for additive background noise, and so cannot be used to explain the change in precision.

Using an identical procedure, we also computed the CV error for the multislice simulations, shown in Figure~\ref{fig:multislice}c and g. Because the lattice fitting in the CV approach uses only half the identified Bragg spots, the CV error is always higher than the RMS least squares residual. The trend is the same as for the RMS error in both thicknesses tested. For the 5~nm model the CV error is 3--7~times higher for the normal probes compared to the bullseyes, while for the 20~nm model the CV error is up to 29~times higher for the normal probe compared to the bullseye probe.  

Strain maps produced from each region of the Si wedge using cross-correlation to locate the diffraction disks are shown in Figure~\ref{fig:experiment}(i)--(l), and the standard deviation of the measured strain in each region is marked on the maps. Strain values are referenced to the median measured lattice in each scan region. In the thin region of the Si wedge, the normal probe registers strains of approximately $\pm 1.5\%$ across the scanned area. Bending in the thin region of the wedge leads to mistilt of a few milliradians across the scan region, which shifts the center of illumination of the pattern and the centers of mass of the diffraction disks. While sample mistilt does change the projected lattice spacing and thus the expected diffraction disk positions very slightly (on the order of 0.1\% for the magnitude of tilting we observed), the strong position bias towards the center of mass of the disks when using normal probes leads to large variation in the measured strain. When using bullseye probes on the same sample region, the strain is measured as only $\pm 0.1\%$. In the thick region of the wedge the sample appears flat across the field of view---here, the normal probes show smoothly varying strain from top-to-bottom of the scan, likely due to variation in sample thickness altering the fine structure inside the diffraction disks. The bullseye probe again reveals a flat strain distribution with standard deviation of about 0.1\%.

Recently, Gizzinati \textit{et al} \cite{guzzinati2019electron} demonstrated a Bessel beam structured probe for improved 4D-STEM strain mapping precision. By inserting an annulus in the second condenser aperture of an aberration corrected microscope, they produce a hollow cone probe with a semi-convergence angle of $6$~mrad, giving spatial resolution of 1.3~nm. In the present work (without aberration correction), we measured the full width at half maximum of the probe to be 2.7~nm at 3~mrad. This follows the expected scaling, where using a semi-convergence angle half as large leads to a doubling of the real-space probe size. Probes with strong amplitude structuring will necessarily sacrifice real space resolution for momentum resolution because of missing frequencies in the wavefunction. In the Bessel beam approach this broadening of the probe is partially mitigated because the beam is so sparse in momentum space that diffracted beams can overlap without substantial interference---this permits high convergence angles, leading to smaller realspace probes. Our approach is practically limited to $\alpha<\theta_B/4$, where $\alpha$ is the semi-convergence angle and $\theta_B$ is the Bragg scattering angle. By observing the variation in strain in a flat region of the sample, they estimated their strain precision as $2.5\times10^{-4}$. In this work, by comparison, we observed strain precision as good as $2.2\times10^{-4}$~--~$2.9\times10^{-5}$ for different simulated models (by the RMS residual metric), and $8\times10^{-3}$~--~$3\times10^{-3}$ in experiments on an Si wedge (by the CV metric).

The patterned probe approach has parallels to Sobel (edge-enhancing) filtering of the diffraction images. In using the Sobel filter, we assume that the CBED disks should have sharp edges and uniform intensity, and so once filtered the disks become rings. Naturally, dynamical structure in the CBED disks will also create edges that are exaggerated by the filter, and indeed Pekin \textit{et al}.\ found that Sobel filtering improves precision for flat disks but causes artifacts when dynamical structure is present \cite{pekin2017strain}. By applying the patterning to the probe before the sample, we avoid this drawback by adding many edges that are defined by the template. 

Compared to other TEM strain mapping techniques, 4D-STEM has generally been reported to have lower precision and lower resolution than other TEM strain mapping techniques, such as atomic resolution imaging and darkfield holography. In particular, 4D-STEM strain mapping is not not possible at atomic resolution as phase interference between scattered beams complicates measurement of the Bragg scattering. However, 4D-STEM offers the greatest flexibility with regards to sample type and orientation, allowing analyses of partially or completely amorphous samples, polycrystals, highly defective materials, and low-symmetry oriented crystals. Simultaneous measurement of other signals is also possible from 4D-STEM data, such as differential phase contrast (DPC) for electric field mapping.  With patterned probes, the precision of 4D-STEM strain measurements can rival that of other techniques, though still with the trade-off between resolution and flexibility. Detailed comparisons of the various TEM strain mapping techniques are available in the literature \cite{COOPER2016145,BECHE201310}.

Early studies on 4D-STEM strain mapping were limited in acquisition speed by CCD detectors, so most of these works used well-exposed diffraction patterns with high electron doses (qualitatively, these patterns match those in our simulations where we find precision saturates and becomes dose-insensitive). With the latest generation of fast detectors operated at full speed, the dose per pattern is limited by the brightness and coherence of the source, limiting the attainable precision.
A potential drawback of the patterned probe approach is the reduction in probe intensity. Our bullseye grating reduces the beam intensity by roughly half, leading to either a twofold increase in exposure time for the same dose (with accompanying increase in sample drift during a measurement) or the use of less coherent illumination to increase the probe current (which will degrade the probe size and quality of the diffraction patterns). However, as the bullseye probes tested here give a roughly fourfold improvement in precision at equal dose while the precision scales as the square root of the dose, higher precision can be realized without compensating for the lost current. For thin samples, where the scaling laws (Eqs.~\ref{eqn:variance-bullseye} \& \ref{eqn:variance2}) hold, using the bullseye aperture with identical microscope settings would give roughly $4/\sqrt{2}\approx2.8$ times improvement, while for thick samples the improvement can be larger.

Using the bullseye patterned probes also requires more pixels per CBED spot in order to resolve the fine pattern features with high fidelity. In some cases, particularly for thick samples and CCD detectors, this requirement necessitates ``spreading out'' the diffracted beam intensity over more pixels, lowering the signal-to-noise ratio. Conversely, when using direct electron detectors with limited dynamic range, the ability to operate at a higher convergence angle and distribute intensity over more pixels can be advantageous.

\section{Conclusion}
\label{Sec:Conc}
We have demonstrated how electron probes with patterning in momentum space can improve the precision of the CBED disk detection procedure used for calculating strain from scanning diffraction data. This approach greatly improves the precision of strain measurements from thicker samples by reducing the systematic errors that arise when locating the Bragg disks in diffraction patterns through thick samples, and potentially enabling more reliable temperature and subtle deformation measurements. In strain maps from a nominally unstrained silicon sample we observe that the anomalous strain measurements caused by dynamical effects are reduced from $\epsilon = \pm1.5\%$ to about $\pm0.1\%$. The specific findings can be summarized as follows:

\begin{itemize}
    \item Imprinting structure on the STEM probe in momentum space adds known, constant contrast to CBED disks which improves the precision of cross-correlation of a known template to the experimental data. For an evenly illuminated CBED disk the position measurement precision increases by a factor determined by the ``roughness'' of the pattern, independent of dose. For the ``bullseye'' pattern we used, a 4-fold improvement is expected.
    
    \item In multislice simulations of a thin sample with largely kinematic scattering, the strain mapping precision improved by a factor of $\approx4$~times at all doses, in agreement with theory. At high doses, the precision reaches a plateau, limited by the subpixel fitting. In simulations of thick samples, where dynamical scattering causes uneven illumination of the CBED disks, the precision improvement is even greater, up to a factor of 29~times.
    
    \item In experiments on an unstrained Si sample, we observe an improvement in precision of about 12 times for both thick and thin regions of the wedge sample. Due to the inelastic background scattering, the hybrid correlation algorithm performs better than cross-correlation when using a normal circular probe. Using the bullseye patterned probe, the cross-correlation algorithm performs best. Strain maps produced from thick and thin regions of the silicon sample show substantially flatter strain across the same sample regions.
\end{itemize}

\section{Acknowledgements}
SEZ was supported by the National Science Foundation under STROBE Grant No. DMR 1548924. BS and LH were supported by the Toyota Research Institute. CO acknowledges
support from the US Department of Energy Early Career Research Program. We would like to thank Shreyas Cholia, Matthew Henderson, Rollin Thomas, and Ludovico Bianchi from the National Energy Research Scientific Computing Center (NERSC) for assistance with computations. NERSC is a U.S. Department of Energy Office of Science User Facility operated under Contract No. DE-AC02-05CH11231. This research used the Savio computational cluster resource provided by the Berkeley Research Computing program at the University of California, Berkeley (supported by the UC Berkeley Chancellor, Vice Chancellor for Research, and Chief Information Officer). Work at the Molecular Foundry was supported by the Office of Science, Office of Basic Energy Sciences, of the U.S. Department of Energy under Contract No. DE-AC02-05CH11231. 

\bibliographystyle{model1-num-names}
\bibliography{4DSTEMrefs.bib}

\begin{thebibliography}{46}
\expandafter\ifx\csname natexlab\endcsname\relax\def\natexlab#1{#1}\fi
\providecommand{\bibinfo}[2]{#2}
\ifx\xfnm\relax \def\xfnm[#1]{\unskip,\space#1}\fi
\bibitem[{Wang et~al.(2012)Wang, Li, Shan, Li, Sun, and Ma}]{wang2012sample}
\bibinfo{author}{Z.-J. Wang}, \bibinfo{author}{Q.-J. Li},
  \bibinfo{author}{Z.-W. Shan}, \bibinfo{author}{J.~Li},
  \bibinfo{author}{J.~Sun}, \bibinfo{author}{E.~Ma},
\newblock \bibinfo{title}{Sample size effects on the large strain bursts in
  submicron aluminum pillars},
\newblock \bibinfo{journal}{Applied Physics Letters} \bibinfo{volume}{100}
  (\bibinfo{year}{2012}) \bibinfo{pages}{071906}.
\bibitem[{Bedell et~al.(2014)Bedell, Khakifirooz, and
  Sadana}]{bedell2014strain}
\bibinfo{author}{S.~Bedell}, \bibinfo{author}{A.~Khakifirooz},
  \bibinfo{author}{D.~Sadana},
\newblock \bibinfo{title}{Strain scaling for {CMOS}},
\newblock \bibinfo{journal}{{MRS} Bulletin} \bibinfo{volume}{39}
  (\bibinfo{year}{2014}) \bibinfo{pages}{131--137}.
\bibitem[{Li et~al.(2014)Li, Shan, and Ma}]{li2014elastic}
\bibinfo{author}{J.~Li}, \bibinfo{author}{Z.~Shan}, \bibinfo{author}{E.~Ma},
\newblock \bibinfo{title}{Elastic strain engineering for unprecedented
  materials properties},
\newblock \bibinfo{journal}{MRS Bulletin} \bibinfo{volume}{39}
  (\bibinfo{year}{2014}) \bibinfo{pages}{108--114}.
\bibitem[{Holt et~al.(2014)Holt, Hruszkewycz, Murray, Holt, Paskiewicz, and
  Fuoss}]{holt2014xray}
\bibinfo{author}{M.~V. Holt}, \bibinfo{author}{S.~O. Hruszkewycz},
  \bibinfo{author}{C.~E. Murray}, \bibinfo{author}{J.~R. Holt},
  \bibinfo{author}{D.~M. Paskiewicz}, \bibinfo{author}{P.~H. Fuoss},
\newblock \bibinfo{title}{Strain imaging of nanoscale semiconductor
  heterostructures with {X}-ray bragg projection ptychography},
\newblock \bibinfo{journal}{Phys. Rev. Lett.} \bibinfo{volume}{112}
  (\bibinfo{year}{2014}) \bibinfo{pages}{165502}.
\bibitem[{Robinson and Harder(2009)}]{robinson2009coherent}
\bibinfo{author}{I.~Robinson}, \bibinfo{author}{R.~Harder},
\newblock \bibinfo{title}{Coherent {X}-ray diffraction imaging of strain at the
  nanoscale},
\newblock \bibinfo{journal}{Nature Materials} \bibinfo{volume}{8}
  (\bibinfo{year}{2009}) \bibinfo{pages}{291}.
\bibitem[{Koch et~al.(2010)Koch, {\"O}zd{\"o}l, and van
  Aken}]{koch2010efficient}
\bibinfo{author}{C.~T. Koch}, \bibinfo{author}{V.~B. {\"O}zd{\"o}l},
  \bibinfo{author}{P.~A. van Aken},
\newblock \bibinfo{title}{An efficient, simple, and precise way to map strain
  with nanometer resolution in semiconductor devices},
\newblock \bibinfo{journal}{Applied Physics Letters} \bibinfo{volume}{96}
  (\bibinfo{year}{2010}) \bibinfo{pages}{091901}.
\bibitem[{Cooper et~al.(2016)Cooper, Denneulin, Bernier, B\'{e}ch\'{e}, and
  Rouvi\`{e}re}]{COOPER2016145}
\bibinfo{author}{D.~Cooper}, \bibinfo{author}{T.~Denneulin},
  \bibinfo{author}{N.~Bernier}, \bibinfo{author}{A.~B\'{e}ch\'{e}},
  \bibinfo{author}{J.-L. Rouvi\`{e}re},
\newblock \bibinfo{title}{Strain mapping of semiconductor specimens with
  nm-scale resolution in a transmission electron microscope},
\newblock \bibinfo{journal}{Micron} \bibinfo{volume}{80} (\bibinfo{year}{2016})
  \bibinfo{pages}{145 -- 165}.
\bibitem[{Bierwolf et~al.(1993)Bierwolf, Hohenstein, Phillipp, Brandt, Crook,
  and Ploog}]{bierwolf1993direct}
\bibinfo{author}{R.~Bierwolf}, \bibinfo{author}{M.~Hohenstein},
  \bibinfo{author}{F.~Phillipp}, \bibinfo{author}{O.~Brandt},
  \bibinfo{author}{G.~Crook}, \bibinfo{author}{K.~Ploog},
\newblock \bibinfo{title}{Direct measurement of local lattice distortions in
  strained layer structures by hrem},
\newblock \bibinfo{journal}{Ultramicroscopy} \bibinfo{volume}{49}
  (\bibinfo{year}{1993}) \bibinfo{pages}{273--285}.
\bibitem[{Galindo et~al.(2007)Galindo, Kret, Sanchez, Laval, Yáñez, Pizarro,
  Guerrero, Ben, and Molina}]{GALINDO20071186}
\bibinfo{author}{P.~L. Galindo}, \bibinfo{author}{S.~Kret},
  \bibinfo{author}{A.~M. Sanchez}, \bibinfo{author}{J.-Y. Laval},
  \bibinfo{author}{A.~Yáñez}, \bibinfo{author}{J.~Pizarro},
  \bibinfo{author}{E.~Guerrero}, \bibinfo{author}{T.~Ben},
  \bibinfo{author}{S.~I. Molina},
\newblock \bibinfo{title}{The peak pairs algorithm for strain mapping from
  hrtem images},
\newblock \bibinfo{journal}{Ultramicroscopy} \bibinfo{volume}{107}
  (\bibinfo{year}{2007}) \bibinfo{pages}{1186 -- 1193}.
\bibitem[{H{\"y}tch et~al.(1998)H{\"y}tch, Snoeck, and Kilaas}]{HYTCH1998131}
\bibinfo{author}{M.~H{\"y}tch}, \bibinfo{author}{E.~Snoeck},
  \bibinfo{author}{R.~Kilaas},
\newblock \bibinfo{title}{Quantitative measurement of displacement and strain
  fields from hrem micrographs},
\newblock \bibinfo{journal}{Ultramicroscopy} \bibinfo{volume}{74}
  (\bibinfo{year}{1998}) \bibinfo{pages}{131 -- 146}.
\bibitem[{Jones et~al.(1977)Jones, Rackham, and Steeds}]{jones1977higher}
\bibinfo{author}{P.~Jones}, \bibinfo{author}{G.~Rackham},
  \bibinfo{author}{J.~W. Steeds},
\newblock \bibinfo{title}{Higher order {L}aue zone effects in electron
  diffraction and their use in lattice parameter determination},
\newblock \bibinfo{journal}{Proceedings of the Royal Society of London A.
  Mathematical and Physical Sciences} \bibinfo{volume}{354}
  (\bibinfo{year}{1977}) \bibinfo{pages}{197--222}.
\bibitem[{Zhang et~al.(2006)Zhang, Istratov, Weber, Kisielowski, He, Nelson,
  and Spence}]{zhang2006direct}
\bibinfo{author}{P.~Zhang}, \bibinfo{author}{A.~A. Istratov},
  \bibinfo{author}{E.~R. Weber}, \bibinfo{author}{C.~Kisielowski},
  \bibinfo{author}{H.~He}, \bibinfo{author}{C.~Nelson}, \bibinfo{author}{J.~C.
  Spence},
\newblock \bibinfo{title}{Direct strain measurement in a 65 nm node strained
  silicon transistor by convergent-beam electron diffraction},
\newblock \bibinfo{journal}{Applied Physics Letters} \bibinfo{volume}{89}
  (\bibinfo{year}{2006}) \bibinfo{pages}{161907}.
\bibitem[{Cl{\'e}ment et~al.(2004)Cl{\'e}ment, Pantel, Kwakman, and
  Rouvi{\`e}re}]{clement2004strain}
\bibinfo{author}{L.~Cl{\'e}ment}, \bibinfo{author}{R.~Pantel},
  \bibinfo{author}{L.~T. Kwakman}, \bibinfo{author}{J.~Rouvi{\`e}re},
\newblock \bibinfo{title}{Strain measurements by convergent-beam electron
  diffraction: The importance of stress relaxation in lamella preparations},
\newblock \bibinfo{journal}{Applied Physics Letters} \bibinfo{volume}{85}
  (\bibinfo{year}{2004}) \bibinfo{pages}{651--653}.
\bibitem[{H{\"y}tch and Minor(2014)}]{hytch2014observing}
\bibinfo{author}{M.~J. H{\"y}tch}, \bibinfo{author}{A.~M. Minor},
\newblock \bibinfo{title}{Observing and measuring strain in nanostructures and
  devices with transmission electron microscopy},
\newblock \bibinfo{journal}{MRS bulletin} \bibinfo{volume}{39}
  (\bibinfo{year}{2014}) \bibinfo{pages}{138--146}.
\bibitem[{Pennycook(2011)}]{pennycook2011_STEM_textbook}
\bibinfo{author}{S.~J. Pennycook},
\newblock \bibinfo{title}{A scan through the history of {STEM}},
\newblock in: \bibinfo{booktitle}{Scanning Transmission Electron Microscopy},
  \bibinfo{publisher}{Springer}, \bibinfo{year}{2011}, pp.
  \bibinfo{pages}{1--90}.
\bibitem[{Ophus(2019)}]{ophus_2019}
\bibinfo{author}{C.~Ophus},
\newblock \bibinfo{title}{Four-dimensional scanning transmission electron
  microscopy {(4D-STEM)}: From scanning nanodiffraction to ptychography and
  beyond},
\newblock \bibinfo{journal}{Microscopy and Microanalysis} \bibinfo{volume}{25}
  (\bibinfo{year}{2019}) \bibinfo{pages}{563–582}.
\bibitem[{Rauch and Dupuy(2005)}]{rauch2005rapid}
\bibinfo{author}{E.~Rauch}, \bibinfo{author}{L.~Dupuy},
\newblock \bibinfo{title}{Rapid spot diffraction patterns identification
  through template matching},
\newblock \bibinfo{journal}{Archives of Metallurgy and Materials}
  \bibinfo{volume}{50} (\bibinfo{year}{2005}) \bibinfo{pages}{87--99}.
\bibitem[{Brunetti et~al.(2011)Brunetti, Robert, Bayle-Guillemaud,
  Rouvi\`{e}re, Rauch, Martin, Colin, Bertin, and
  Cayron}]{brunetti2011confirmation}
\bibinfo{author}{G.~Brunetti}, \bibinfo{author}{D.~Robert},
  \bibinfo{author}{P.~Bayle-Guillemaud}, \bibinfo{author}{J.~Rouvi\`{e}re},
  \bibinfo{author}{E.~Rauch}, \bibinfo{author}{J.~Martin},
  \bibinfo{author}{J.~Colin}, \bibinfo{author}{F.~Bertin},
  \bibinfo{author}{C.~Cayron},
\newblock \bibinfo{title}{Confirmation of the domino-cascade model by
  {LiFePO$_4$/FePO$_4$} precession electron diffraction},
\newblock \bibinfo{journal}{Chemistry of Materials} \bibinfo{volume}{23}
  (\bibinfo{year}{2011}) \bibinfo{pages}{4515--4524}.
\bibitem[{Panova et~al.(2016)Panova, Chen, Bustillo, Ophus, Bhatt, Balsara, and
  Minor}]{panova2016orientation}
\bibinfo{author}{O.~Panova}, \bibinfo{author}{X.~C. Chen},
  \bibinfo{author}{K.~C. Bustillo}, \bibinfo{author}{C.~Ophus},
  \bibinfo{author}{M.~P. Bhatt}, \bibinfo{author}{N.~Balsara},
  \bibinfo{author}{A.~M. Minor},
\newblock \bibinfo{title}{Orientation mapping of semicrystalline polymers using
  scanning electron nanobeam diffraction},
\newblock \bibinfo{journal}{Micron} \bibinfo{volume}{88} (\bibinfo{year}{2016})
  \bibinfo{pages}{30--36}.
\bibitem[{Liu et~al.(2015)Liu, Lumpkin, Petersen, Etheridge, and
  Bourgeois}]{liu2015interpretation}
\bibinfo{author}{A.~C. Liu}, \bibinfo{author}{G.~R. Lumpkin},
  \bibinfo{author}{T.~C. Petersen}, \bibinfo{author}{J.~Etheridge},
  \bibinfo{author}{L.~Bourgeois},
\newblock \bibinfo{title}{Interpretation of angular symmetries in electron
  nanodiffraction patterns from thin amorphous specimens},
\newblock \bibinfo{journal}{Acta Crystallographica Section A: Foundations and
  Advances} \bibinfo{volume}{71} (\bibinfo{year}{2015})
  \bibinfo{pages}{473--482}.
\bibitem[{LeBeau et~al.(2010)LeBeau, Findlay, Allen, and
  Stemmer}]{lebeau2010position}
\bibinfo{author}{J.~M. LeBeau}, \bibinfo{author}{S.~D. Findlay},
  \bibinfo{author}{L.~J. Allen}, \bibinfo{author}{S.~Stemmer},
\newblock \bibinfo{title}{Position averaged convergent beam electron
  diffraction: Theory and applications},
\newblock \bibinfo{journal}{Ultramicroscopy} \bibinfo{volume}{110}
  (\bibinfo{year}{2010}) \bibinfo{pages}{118--125}.
\bibitem[{Zhu et~al.(2012)Zhu, Radtke, and Botton}]{zhu2012bonding}
\bibinfo{author}{G.-Z. Zhu}, \bibinfo{author}{G.~Radtke},
  \bibinfo{author}{G.~A. Botton},
\newblock \bibinfo{title}{Bonding and structure of a reconstructed (001)
  surface of {SrTiO}$_3$ from {TEM}},
\newblock \bibinfo{journal}{Nature} \bibinfo{volume}{490}
  (\bibinfo{year}{2012}) \bibinfo{pages}{384}.
\bibitem[{Usuda et~al.(2004)Usuda, Mizuno, Tezuka, Sugiyama, Moriyama,
  Nakaharai, and Takagi}]{usuda2004strain}
\bibinfo{author}{K.~Usuda}, \bibinfo{author}{T.~Mizuno},
  \bibinfo{author}{T.~Tezuka}, \bibinfo{author}{N.~Sugiyama},
  \bibinfo{author}{Y.~Moriyama}, \bibinfo{author}{S.~Nakaharai},
  \bibinfo{author}{S.~Takagi},
\newblock \bibinfo{title}{Strain relaxation of strained-{Si} layers on
  {SiGe}-on-insulator ({SGOI}) structures after mesa isolation},
\newblock \bibinfo{journal}{Applied Surface Science} \bibinfo{volume}{224}
  (\bibinfo{year}{2004}) \bibinfo{pages}{113--116}.
\bibitem[{Pekin et~al.(2017)Pekin, Gammer, Ciston, Minor, and
  Ophus}]{pekin2017strain}
\bibinfo{author}{T.~C. Pekin}, \bibinfo{author}{C.~Gammer},
  \bibinfo{author}{J.~Ciston}, \bibinfo{author}{A.~M. Minor},
  \bibinfo{author}{C.~Ophus},
\newblock \bibinfo{title}{Optimizing disk registration algorithms for nanobeam
  electron diffraction strain mapping},
\newblock \bibinfo{journal}{Ultramicroscopy} \bibinfo{volume}{176}
  (\bibinfo{year}{2017}) \bibinfo{pages}{170--176}.
\bibitem[{Gammer et~al.(2018)Gammer, Ophus, Pekin, Eckert, and
  Minor}]{gammer2018local}
\bibinfo{author}{C.~Gammer}, \bibinfo{author}{C.~Ophus}, \bibinfo{author}{T.~C.
  Pekin}, \bibinfo{author}{J.~Eckert}, \bibinfo{author}{A.~M. Minor},
\newblock \bibinfo{title}{Local nanoscale strain mapping of a metallic glass
  during \textit{in situ} testing},
\newblock \bibinfo{journal}{Applied Physics Letters} \bibinfo{volume}{112}
  (\bibinfo{year}{2018}) \bibinfo{pages}{171905}.
\bibitem[{Pekin et~al.(2018)Pekin, Gammer, Ciston, Ophus, and
  Minor}]{pekin2018situ}
\bibinfo{author}{T.~C. Pekin}, \bibinfo{author}{C.~Gammer},
  \bibinfo{author}{J.~Ciston}, \bibinfo{author}{C.~Ophus},
  \bibinfo{author}{A.~M. Minor},
\newblock \bibinfo{title}{In situ nanobeam electron diffraction strain mapping
  of planar slip in stainless steel},
\newblock \bibinfo{journal}{Scripta Materialia} \bibinfo{volume}{146}
  (\bibinfo{year}{2018}) \bibinfo{pages}{87--90}.
\bibitem[{Han et~al.(2018)Han, Nguyen, Cao, Cueva, Xie, Tate, Purohit, Gruner,
  Park, and Muller}]{han2018strain}
\bibinfo{author}{Y.~Han}, \bibinfo{author}{K.~Nguyen},
  \bibinfo{author}{M.~Cao}, \bibinfo{author}{P.~Cueva},
  \bibinfo{author}{S.~Xie}, \bibinfo{author}{M.~Tate},
  \bibinfo{author}{P.~Purohit}, \bibinfo{author}{S.~Gruner},
  \bibinfo{author}{J.~Park}, \bibinfo{author}{D.~Muller},
\newblock \bibinfo{title}{Strain mapping of two-dimensional heterostructures
  with subpicometer precision.},
\newblock \bibinfo{journal}{Nano Letters} \bibinfo{volume}{18}
  (\bibinfo{year}{2018}) \bibinfo{pages}{3746}.
\bibitem[{B{\'e}ch{\'e} et~al.(2009)B{\'e}ch{\'e}, Rouvi\`{e}re, Cl{\'e}ment,
  and Hartmann}]{beche2009improved}
\bibinfo{author}{A.~B{\'e}ch{\'e}}, \bibinfo{author}{J.~Rouvi\`{e}re},
  \bibinfo{author}{L.~Cl{\'e}ment}, \bibinfo{author}{J.~Hartmann},
\newblock \bibinfo{title}{Improved precision in strain measurement using
  nanobeam electron diffraction},
\newblock \bibinfo{journal}{Applied Physics Letters} \bibinfo{volume}{95}
  (\bibinfo{year}{2009}) \bibinfo{pages}{123114}.
\bibitem[{Rouvi\`{e}re(2013)}]{rouviere2013method}
\bibinfo{author}{J.-L. Rouvi\`{e}re}, \bibinfo{title}{Method to facilitate
  positioning of diffraction spots}, \bibinfo{year}{2013}. \bibinfo{note}{US
  Patent App. 13/877,904}.
\bibitem[{Guzzinati et~al.(2019)Guzzinati, Ghielens, Mahr, B{\'e}ch{\'e},
  Rosenauer, Calders, and Verbeeck}]{guzzinati2019electron}
\bibinfo{author}{G.~Guzzinati}, \bibinfo{author}{W.~Ghielens},
  \bibinfo{author}{C.~Mahr}, \bibinfo{author}{A.~B{\'e}ch{\'e}},
  \bibinfo{author}{A.~Rosenauer}, \bibinfo{author}{T.~Calders},
  \bibinfo{author}{J.~Verbeeck},
\newblock \bibinfo{title}{Electron bessel beam diffraction for precise and
  accurate nanoscale strain mapping},
\newblock \bibinfo{journal}{arXiv:1902.06979}  (\bibinfo{year}{2019}).
\bibitem[{Mahr et~al.(2015)Mahr, M{\"u}ller-Caspary, Grieb, Schowalter,
  Mehrtens, Krause, Zillmann, and Rosenauer}]{mahr2015theoretical}
\bibinfo{author}{C.~Mahr}, \bibinfo{author}{K.~M{\"u}ller-Caspary},
  \bibinfo{author}{T.~Grieb}, \bibinfo{author}{M.~Schowalter},
  \bibinfo{author}{T.~Mehrtens}, \bibinfo{author}{F.~F. Krause},
  \bibinfo{author}{D.~Zillmann}, \bibinfo{author}{A.~Rosenauer},
\newblock \bibinfo{title}{Theoretical study of precision and accuracy of strain
  analysis by nano-beam electron diffraction},
\newblock \bibinfo{journal}{Ultramicroscopy} \bibinfo{volume}{158}
  (\bibinfo{year}{2015}) \bibinfo{pages}{38--48}.
\bibitem[{Grieb et~al.(2017)Grieb, Krause, Mahr, Zillmann, M{\"u}ller-Caspary,
  Schowalter, and Rosenauer}]{grieb2017optimization}
\bibinfo{author}{T.~Grieb}, \bibinfo{author}{F.~F. Krause},
  \bibinfo{author}{C.~Mahr}, \bibinfo{author}{D.~Zillmann},
  \bibinfo{author}{K.~M{\"u}ller-Caspary}, \bibinfo{author}{M.~Schowalter},
  \bibinfo{author}{A.~Rosenauer},
\newblock \bibinfo{title}{Optimization of {NBED} simulations for disc-detection
  measurements},
\newblock \bibinfo{journal}{Ultramicroscopy} \bibinfo{volume}{181}
  (\bibinfo{year}{2017}) \bibinfo{pages}{50--60}.
\bibitem[{Cowley and Moodie(1957)}]{cowley1957scattering}
\bibinfo{author}{J.~M. Cowley}, \bibinfo{author}{A.~F. Moodie},
\newblock \bibinfo{title}{The scattering of electrons by atoms and crystals.
  {I}. a new theoretical approach},
\newblock \bibinfo{journal}{Acta Crystallographica} \bibinfo{volume}{10}
  (\bibinfo{year}{1957}) \bibinfo{pages}{609--619}.
\bibitem[{Rouvi\`{e}re et~al.(2013)Rouvi\`{e}re, B{\'e}ch{\'e}, Martin,
  Denneulin, and Cooper}]{rouviere2013improved}
\bibinfo{author}{J.-L. Rouvi\`{e}re}, \bibinfo{author}{A.~B{\'e}ch{\'e}},
  \bibinfo{author}{Y.~Martin}, \bibinfo{author}{T.~Denneulin},
  \bibinfo{author}{D.~Cooper},
\newblock \bibinfo{title}{Improved strain precision with high spatial
  resolution using nanobeam precession electron diffraction},
\newblock \bibinfo{journal}{Applied Physics Letters} \bibinfo{volume}{103}
  (\bibinfo{year}{2013}) \bibinfo{pages}{241913}.
\bibitem[{M{\"u}ller et~al.(2012)M{\"u}ller, Ryll, Ordavo, Ihle, Str{\"u}der,
  Volz, Zweck, Soltau, and Rosenauer}]{muller2012scanning}
\bibinfo{author}{K.~M{\"u}ller}, \bibinfo{author}{H.~Ryll},
  \bibinfo{author}{I.~Ordavo}, \bibinfo{author}{S.~Ihle},
  \bibinfo{author}{L.~Str{\"u}der}, \bibinfo{author}{K.~Volz},
  \bibinfo{author}{J.~Zweck}, \bibinfo{author}{H.~Soltau},
  \bibinfo{author}{A.~Rosenauer},
\newblock \bibinfo{title}{Scanning transmission electron microscopy strain
  measurement from millisecond frames of a direct electron charge coupled
  device},
\newblock \bibinfo{journal}{Applied Physics Letters} \bibinfo{volume}{101}
  (\bibinfo{year}{2012}) \bibinfo{pages}{212110}.
\bibitem[{H{\"a}hnel et~al.(2012)H{\"a}hnel, Reiche, Moutanabbir, Blumtritt,
  Geisler, H{\"o}ntschel, and Engelmann}]{hahnel2012improving}
\bibinfo{author}{A.~H{\"a}hnel}, \bibinfo{author}{M.~Reiche},
  \bibinfo{author}{O.~Moutanabbir}, \bibinfo{author}{H.~Blumtritt},
  \bibinfo{author}{H.~Geisler}, \bibinfo{author}{J.~H{\"o}ntschel},
  \bibinfo{author}{H.-J. Engelmann},
\newblock \bibinfo{title}{Improving accuracy and precision of strain analysis
  by energy-filtered nanobeam electron diffraction},
\newblock \bibinfo{journal}{Microscopy and Microanalysis} \bibinfo{volume}{18}
  (\bibinfo{year}{2012}) \bibinfo{pages}{229--240}.
\bibitem[{Wehmeyer et~al.(2018)Wehmeyer, Bustillo, Minor, and
  Dames}]{wehmeyer2018measuring}
\bibinfo{author}{G.~Wehmeyer}, \bibinfo{author}{K.~C. Bustillo},
  \bibinfo{author}{A.~M. Minor}, \bibinfo{author}{C.~Dames},
\newblock \bibinfo{title}{Measuring temperature-dependent thermal diffuse
  scattering using scanning transmission electron microscopy},
\newblock \bibinfo{journal}{Applied Physics Letters} \bibinfo{volume}{113}
  (\bibinfo{year}{2018}) \bibinfo{pages}{253101}.
\bibitem[{Clement et~al.(2018)Clement, Bierbaum, and Sethna}]{clement2018image}
\bibinfo{author}{C.~B. Clement}, \bibinfo{author}{M.~Bierbaum},
  \bibinfo{author}{J.~P. Sethna},
\newblock \bibinfo{title}{Image registration and super resolution from first
  principles},
\newblock \bibinfo{journal}{arXiv:1809.05583}  (\bibinfo{year}{2018}).
\bibitem[{Kirkland(2010)}]{kirkland2010advanced}
\bibinfo{author}{E.~J. Kirkland}, \bibinfo{title}{Advanced computing in
  electron microscopy}, \bibinfo{publisher}{Springer Science \& Business
  Media}, \bibinfo{year}{2010}.
\bibitem[{Ophus(2017)}]{ophus2017fastSim}
\bibinfo{author}{C.~Ophus},
\newblock \bibinfo{title}{A fast image simulation algorithm for scanning
  transmission electron microscopy},
\newblock \bibinfo{journal}{Advanced Structural and Chemical Imaging}
  \bibinfo{volume}{3} (\bibinfo{year}{2017}) \bibinfo{pages}{13}.
\bibitem[{Savitzky et~al.(2019)Savitzky, Zeltmann, Barnard, Dacosta, Brown,
  Henderson, and Ginsburg}]{savitzky}
\bibinfo{author}{B.~Savitzky}, \bibinfo{author}{S.~Zeltmann},
  \bibinfo{author}{E.~Barnard}, \bibinfo{author}{L.~R. Dacosta},
  \bibinfo{author}{H.~G. Brown}, \bibinfo{author}{M.~Henderson},
  \bibinfo{author}{D.~Ginsburg}, \bibinfo{title}{py4dstem: Open source
  processing and analysis of {4D-STEM} data}, \bibinfo{year}{2019}.
\bibitem[{Guizar-Sicairos et~al.(2008)Guizar-Sicairos, Thurman, and
  Fienup}]{Guizar-Sicairos:08}
\bibinfo{author}{M.~Guizar-Sicairos}, \bibinfo{author}{S.~T. Thurman},
  \bibinfo{author}{J.~R. Fienup},
\newblock \bibinfo{title}{Efficient subpixel image registration algorithms},
\newblock \bibinfo{journal}{Opt. Lett.} \bibinfo{volume}{33}
  (\bibinfo{year}{2008}) \bibinfo{pages}{156--158}.
\bibitem[{Soummer et~al.(2007)Soummer, Pueyo, Sivaramakrishnan, and
  Vanderbei}]{Soummer:07}
\bibinfo{author}{R.~Soummer}, \bibinfo{author}{L.~Pueyo},
  \bibinfo{author}{A.~Sivaramakrishnan}, \bibinfo{author}{R.~J. Vanderbei},
\newblock \bibinfo{title}{Fast computation of {L}yot-style coronagraph
  propagation},
\newblock \bibinfo{journal}{Opt. Express} \bibinfo{volume}{15}
  (\bibinfo{year}{2007}) \bibinfo{pages}{15935--15951}.
\bibitem[{Gleason et~al.(1991)Gleason, Hunt, and Jatko}]{gleason1991subpixel}
\bibinfo{author}{S.~S. Gleason}, \bibinfo{author}{M.~A. Hunt},
  \bibinfo{author}{W.~B. Jatko},
\newblock \bibinfo{title}{Subpixel measurement of image features based on
  paraboloid surface fit},
\newblock in: \bibinfo{booktitle}{Machine vision systems integration in
  industry}, volume \bibinfo{volume}{1386},
  \bibinfo{organization}{International Society for Optics and Photonics}, pp.
  \bibinfo{pages}{135--145}.
\bibitem[{Arlot and Celisse(2010)}]{arlot2010survey}
\bibinfo{author}{S.~Arlot}, \bibinfo{author}{A.~Celisse},
\newblock \bibinfo{title}{A survey of cross-validation procedures for model
  selection},
\newblock \bibinfo{journal}{Statistics Surveys} \bibinfo{volume}{4}
  (\bibinfo{year}{2010}) \bibinfo{pages}{40--79}.
\bibitem[{B\'{e}ch\'{e} et~al.(2013)B\'{e}ch\'{e}, Rouvi\`{e}re, Barnes, and
  Cooper}]{BECHE201310}
\bibinfo{author}{A.~B\'{e}ch\'{e}}, \bibinfo{author}{J.~Rouvi\`{e}re},
  \bibinfo{author}{J.~Barnes}, \bibinfo{author}{D.~Cooper},
\newblock \bibinfo{title}{Strain measurement at the nanoscale: Comparison
  between convergent beam electron diffraction, nano-beam electron diffraction,
  high resolution imaging and dark field electron holography},
\newblock \bibinfo{journal}{Ultramicroscopy} \bibinfo{volume}{131}
  (\bibinfo{year}{2013}) \bibinfo{pages}{10 -- 23}.

\end{thebibliography}

\end{document}